\documentclass[preprint,showpacs,preprintnumbers,amsmath,amssymb,floats]{revtex4-1}
\usepackage{blindtext}
\usepackage{graphicx}
\usepackage[english]{babel}
\usepackage{amsmath}
\usepackage{amssymb}
\usepackage{color}
\newcommand{\be}{\begin{eqnarray}}
\newcommand{\ee}{\end{eqnarray}}
\usepackage[toc,page]{appendix}

\begin{document}

\title{ Linear in temperature resistivity and associated mysteries \\ including high temperature superconductivity}

\author{Chandra M. Varma$^*$}
\affiliation{Physics Department, 
University of California, Berkeley, CA. 94704}

\begin{abstract}
Recent experimental results: (i) the measurement of the $T \ln T$ specific heat in cuprates and the
earlier such results in some heavy fermion compounds, (ii) the measurement of the single-particle
scattering rates, (iii) the density fluctuation spectrum in cuprates and (iv) the long standing
results on the linear temperature dependence of the resistivity, show that a 
 theory of 
the quantum-criticality in these compounds based on the solution 
of the dissipative 2D - XY model 
 gives the temperature and frequency dependence of each of them, 
 and the magnitudes of all four with
one dimensionless coupling parameter.
These low frequency or temperature dependences persist to an upper cut-off  which 
is measured to be about the same from the singularity in the specific heat or the saturation 
of the single-particle 
self-energy. The same two parameters are deduced in the analysis 
of results of photoemission experiments to give 
d-wave superconductivity and its transition temperature. The coupling parameter and the
cut-off had been estimated in the microscopic
theory to within a factor of 2.
The simplicity of the results depends on the  discovery that 
orthogonal topological excitations in space and in time 
determine the fluctuations near criticality such that the space and
 time metrics are free of each other. 
The interacting fermions then form a marginal Fermi-liquid.
\end{abstract}
\pacs{}
\date\today
\maketitle
*Visiting Professor

\section{Introduction}
Innumerable papers in the last two decades point out 
that the linear in $T$ resistivity extending to asymptotic low temperatures
 in cuprates, some of the Fe based compounds and some heavy fermion compounds is 
  a great mystery and the most important unsolved 
  problem in condensed matter physics. It was realized very early that fundamental new principles
  must be involved in understanding these and various other associated observed properties related to the breakdown
  of the quasi-particle concept,  and tentative directions of future research laid out 
  \cite{ANDERSON1987}, \cite{CMV-MFL}. Not surprisingly, given the enormous attention devoted to this problem
  by physicists from a variety of different backgrounds, other novel points of views have also developed. 
  These include a branch of string theory, physics of black holes and applications of theory of quantum chaos; 
  see for example the book \cite{Zaanen-book} in which these are summarized.
   
The aim of this brief review is to compare detailed quantitative predictions of a
 theory with 
 a variety of different experiments so that a solution of these remarkable problems may be said to
have been achieved. 
 The solution is quite subtle and relies on quantum-criticality
governed by topological excitations. The answers are unusual, but typical of most subtle problems, extra-ordinarily simple. 
 The model solved is the
  dissipative 2D quantum XY model and the coupling to fermions of its fluctuations. The model 
 has been
 solved by renormalization group methods \cite{Hou-CMV-RG} as accurately as the
 Kosterlitz's solution of the classical XY model \cite{Kosterlitz}, and checked in detail
 by quantum Monte-carlo calculations with some additional results \cite{ZhuChenCMV2015, ZhuHouV2016}. 
 The applicability of the model to
the quantum-criticality of the cuprates  and to anisotropic antiferromagnets
 has also been discussed \cite{Varma2015afmqc} \cite{SchroderZhuV2015}. A brief review of the theory may be found in
  \cite{Varma_IOPrev2016}. The emphasis in this colloque is on
  a quantitative comparison of the predictions with experiments, which was not possible earlier because, even on this prolifically 
  worked at problem, some crucial experimental results have been available only in the past two years. Only the motivation 
  for the
   direction of pursuit of the theory and its principal results are summarized.

Some essential aspects of the fluctuation spectra,
based on a close reading of a variety of
 experiments  
were suggested much earlier \cite{CMV-MFL},
before the microscopic basis was understood and an appropriate theoretical
framework for deriving the results formulated. Now that the
foundations of the unusual  criticality have been found, many important aspects are changed.  
But one of the central results, that the fermions form a {\it marginal Fermi-liquid}, 
which followed from the assumed quantum-critical spectra \cite{CMV-MFL} remains unchanged.
 
Related to the physics of the normal state anomalies is the aspect of the theory giving a
 quantitative
theory of the d-wave superconductivity in these compounds \cite{ASV2010}. As will be 
discussed
below, given the angle dependence of the single-particle scattering rate which is quite unlike
the hot-spot dominated scattering of conventional antiferromagnetic fluctuations, 
 d-wave superconductivity can occur only if the coupling of the critical fluctuations to fermions 
has a unique signature given by the theory. Moreover, the same two
parameters with which various normal state properties are fitted are also deduced
from analysis of angle-resolved photoemission data 
 in a cuprate \cite{Bok_ScienceADV}. 

For the Fe-based compounds and the heavy-fermions, 
 as wide a variety of relevant experimental results as in the cuprates are not yet available. 
 But one can argue from what are available  that the same principles are
at work. This raises an important unresolved issue which will be briefly described. For the case of the cuprates,
the theory of criticality  rests on a elusive proposed order breaking time-reversal and inversion in the under-doped
phase which abuts the quantum-critical region in the phase diagram. The phase transition to such a broken symmetry
has been observed in a wide variety of experiments briefly summarized in Sec. III. But some important properties, the "Fermi-arcs" and magneto-oscillations
with a very small Fermi-surface remain unexplained in this phase. An extension of the broken symmetry to a large period
phase has been suggested \cite{Varma-Per-Order-2019} to explain these very recently but this has not yet been tested in proposed experiments. Only when this
extension is verified and such properties explained can one claim to have a complete theory of the cuprates.

This colloque is organized as follows:
In Sec. II, the four classes of normal state experiments mentioned and the deduction of 
the crucial two parameters determining their magnitude and of d-wave superconductivity 
are summarized. =
In Sec. III, the results of the theory are briefly summarized
to show how the frequency and temperature dependence in each of the experiments
is obtained and why two parameters describe all of them quantitatively. In the concluding section, 
the important unresolved 
problems are mentioned.  

\section{Experimental Results}

\begin{figure}
\includegraphics[width=1.0\columnwidth]{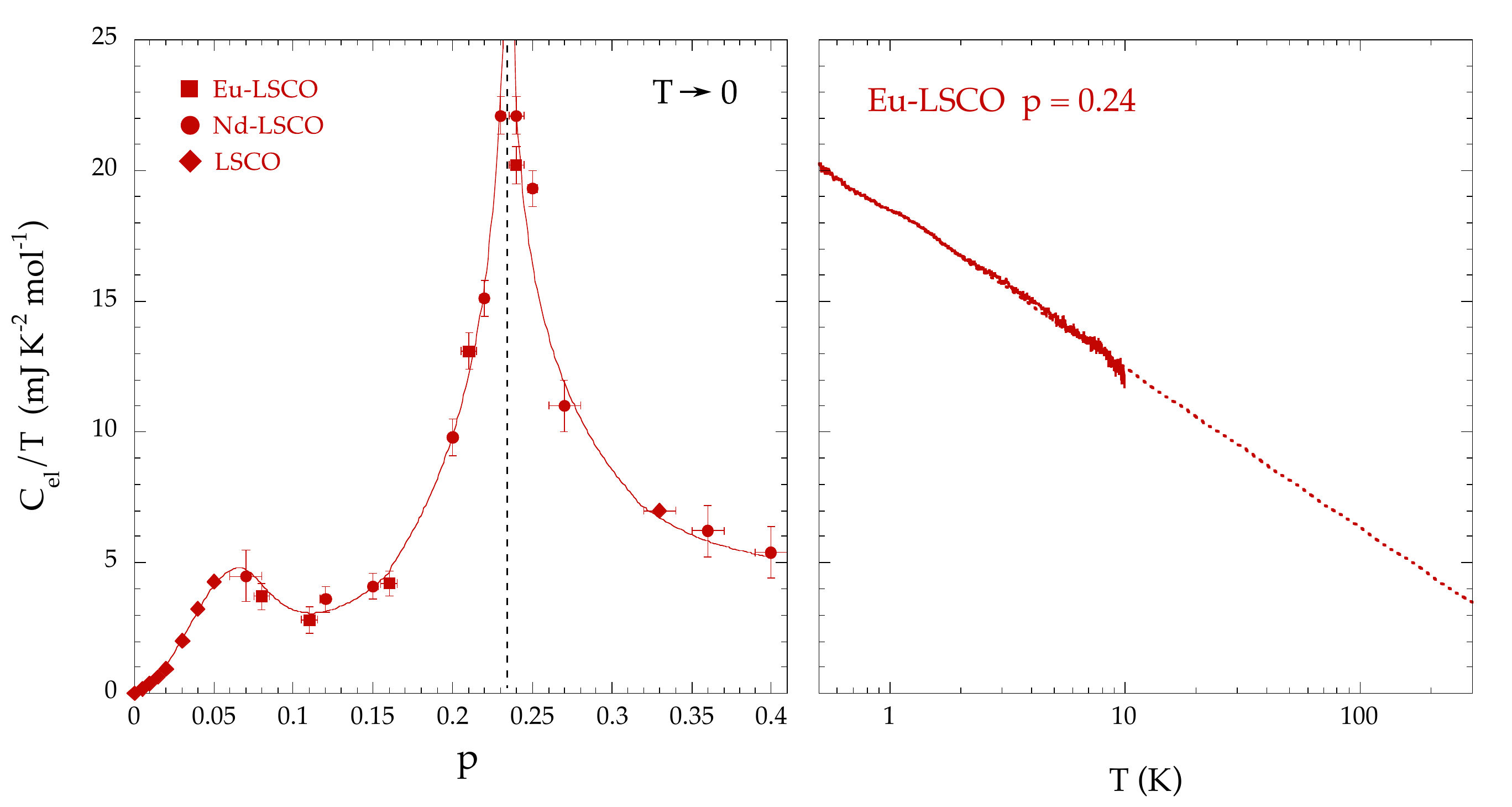}
\caption{The electronic Specific heat  in $La_{2-p}A_pCuO_4$, taken from Fig. S10 of Ref. \cite{Tailleferspht}. 
The left panel shows the data at 0.5 K, the lowest temperature measured 
(in a magnetic field to remove superconductivity). The right panel shows the temperature dependence of
the specific heat nearest to the critical composition $p \approx p_c$. The extraction of the electronic component
from the total specific heat is discussed in \cite{Tailleferspht}} 
\label{fig:QcrSpHt}
\end{figure}

\subsection{Specific heat near quantum-criticality}

 An antiferromagnetic quantum critical point
 for the heavy-fermions and the
Fe-based compounds is evident and there is no question that  the anomalous properties occur associated with it. In the hole-doped cuprates, 
the antiferromagnetic correlation 
length in the region of the linear in T resistivity and other anomalous properties is only of 
order a lattice constant \cite{Bourges-Balatsky-AFM}. A phase diagram
with a quantum critical point in this region (associated with quite a different phase transition) was 
 proposed for the cuprates \cite{cmv1997} with a line of transitions at the onset of the so-called pseudogap phase.
Such a line is now widely accepted. However the clearest thermodynamic evidence for quantum
  criticality was discovered only 
 last year through
 the measurement of the
 the specific singularity in the measured specific heat \cite{Tailleferspht} of the algebraic form predicted in 1989 \cite{CMV-MFL}. 
 Fig. (\ref{fig:QcrSpHt}) from Ref. \cite{Tailleferspht}
 presents both the singularity in the specific heat very close to the critical point in a 
 measurement down to 0.5 K together with 
 the crossover in the singularity on either side of the quantum-critical point. This measurement was possible
 because in the compounds measured, $La_{2-p}A_pCuO_4$ with $A= Nd$ or $Eu$, $T_c$ is low enough to be
 suppressed with fields of about 15 Tesla. 
 
 Very close to
 quantum-criticality, the electronic specific heat fits
 \be
 \label{gamma}
 \frac{C_{el}}{k_BT} (p_c) = \gamma\Big( 1 + \overline{g} \ln \Big(\frac{\overline{T_x}}{T}\Big)\Big).
 \ee
 The logarithmic enhancement of the specific heat is  equivalent to the basic postulates of a marginal
 Fermi-liquid \cite{CMV-MFL} - that the quasi-particle residue go to zero at the critical point as
 \be
 \label{z}
 z_{\hat{p}}(\omega,T) = \frac{1}{1+ g_{\hat{p}} \ln \frac{\pi T_{x{\hat{p}}}}{x}}, ~~ x = max(\pi T, \omega).
 \ee
Following the summary of the theory in Sec. III below, I have assumed that both the coupling constant $g$ and the cut-off
 $T_x$ may have weak dependence on the direction of the momentum ${\bf p}$ at the Fermi-surface. The experimental
 $\overline{g}$ and $\overline{T_x}$ in the specific heat may be taken as the averages of the parameters
  in $z_{\hat{p}}$.
 
 What is plotted in Fig. (\ref{fig:QcrSpHt}) is not the total specific heat divided by $T$, 
 but $C_{el}/T$ obtained by subtracting 
 from the total specific heat
 at a given $p$, all but an observed constant (electronic or Fermi-liquid) contribution 
 to the total specific heat $C_v/T$  at 
 $p=0.16$.  Both are measured at a magnetic field of 8 Tesla to eliminate 
 superconductivity. 
 This serves to eliminate the nuclear Schottky contribution and 
 the  phonon contribution. Using $\gamma \approx 5 mJoules/mole K^2$ at $p=0.24$,
   we may read $\overline{g} $ and the 
  cut-off $\overline{T_x}$,
  from the slope and the intercept by extending the dashed red-line to $0$ in the right side of the figure to be:
 \be
 \label{g,Tx}
 \overline{g} \approx  0.4 \pm 0.1,~~\overline{T_x} \approx~~1,200 \pm 300 K.
 \ee
 The error bars come from the large region over which an extrapolation is 
 necessary to deduce $T_x$ and the (smaller) uncertainty in $\gamma$.
 
 From Fig. (\ref{fig:QcrSpHt} ), one can also deduce the crossover temperature $\xi_T^{-1} (p-p_c)$ to a Fermi-liquid
 \be
 \label{Txx}
  \frac{C_{el}}{T} (p) &=& \gamma (1 + \overline{g} \ln(T_x/\sqrt{T^2 + \xi_T^{-2}(p)}), \\
  \frac{(\xi_T)^{-1}} {T_x} &\propto & \big(\frac{p-p_c}{p_c}\big)^{-\zeta}
 \ee
Given the error bars, $\zeta$ cannot be determined too accurately.
Assuming that the background specific heat coefficient $\gamma$ is independent of T for range of 
$p$ between 0.24 and 0.35, $\zeta \approx 0.5$ is estimated. The calculations summarized in Sec. III
do give this value.

The $T \ln T$ contribution to the specific heat in one heavy Fermion compound in the region of its
linear in $T$ resistivity will be quantitatively discussed later. It would be quite helpful if specific heat is 
measured in the quantum-critical region in other cuprates with low $T_c$ such as Bi2201. In the compounds measured
as well as in others, there is need to measure several close by dopings to the critical point to evaluate the 
cross-over regime quantitatively. The effect of the magnetic field on the criticality also needs to be studied carefully.

\subsection{Single-particle Relaxation rate}

\begin{figure}
\includegraphics[width=1.0\columnwidth]{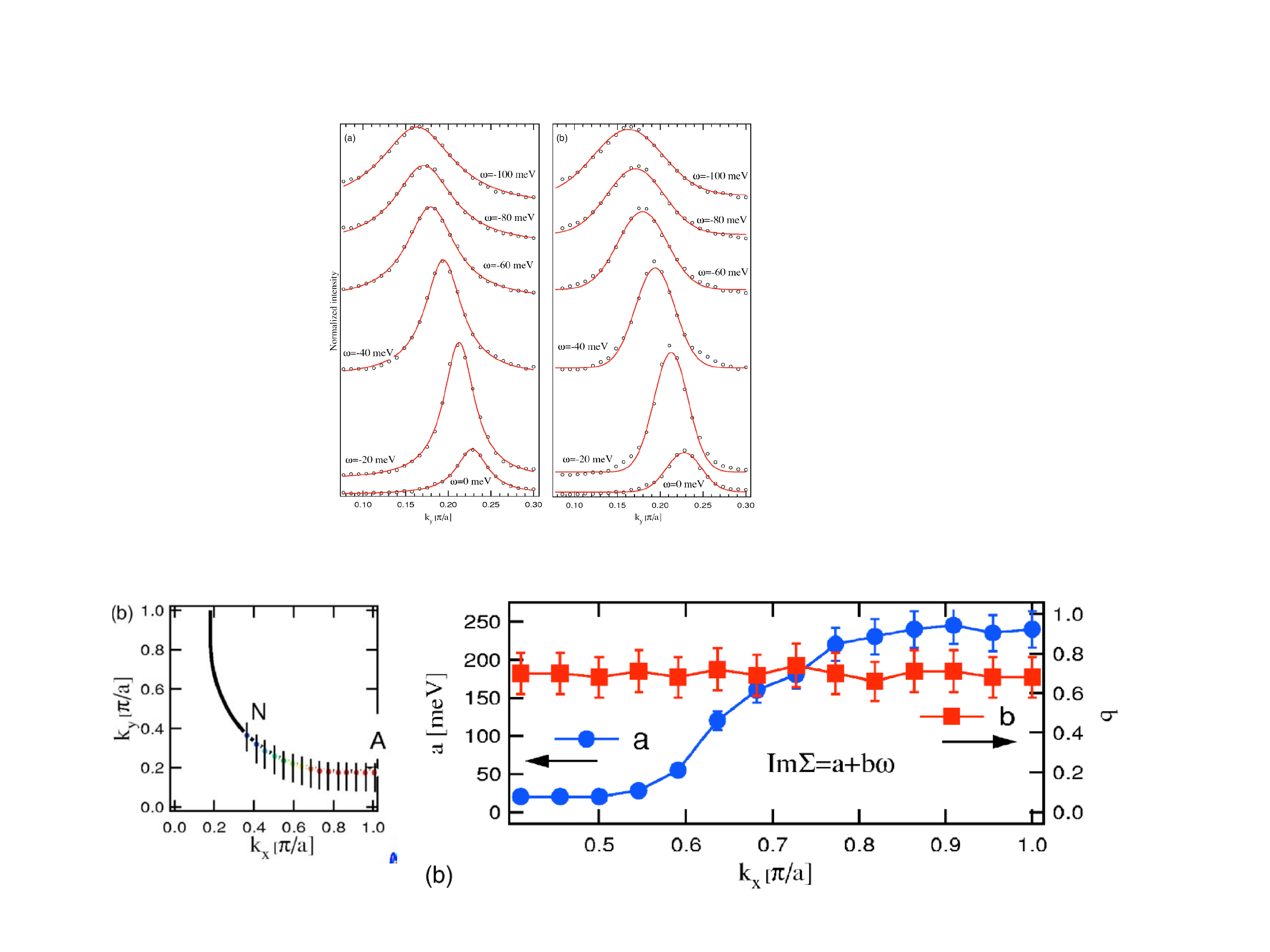}
\caption{Single-particle scattering rate measured by ARPES taken from \cite{KaminskiPR2005}. The top part shows the momentum distribution curves at various energies with a fit with which the parameters of the
self energy are extracted. The data is fit by red-curves which are Lorentzians, proving that the scattering rate is independent of 
momentum perpendicular to the Fermi-surface. The bottom panel on the left shows the points on the Fermi-surface and the directions in which the data was taken.  The bottom panel on the right shows the extracted parameters for the self-energy}
 \label{fig:ImSigma}
\end{figure}

Inelastic single-particle relaxation rate began to be reliably measured in the year 2000 \cite{VallaPRL2000} 
and showed a relaxation rate
proportional to $T$ for $T$ much larger than $\omega$ and proportional to $\omega$ in the
 opposite limit and with evidence that it is nearly independent of momentum both perpendicular to the
 Fermi-surface and along the Fermi-surface. The most complete such measurements 
 arrived in 2005 \cite{KaminskiPR2005}.
 I show the relaxation rate in different directions on the Fermi-surface for $\omega >> T$ in 
 Fig. (\ref{fig:ImSigma}) in that paper. Later measurements have shown a relaxation rate about 20\% smaller
 \cite{BokPRB2010}. Similar results have been obtained by other investigators - for a review, see \cite{ARPES-revZX}. 
 The deduction of the single-particle relaxation rate as a function of frequency, or what is the
 same thing, the imaginary part of the single particle self-energy $Im \Sigma({\bf p}, \omega)$ reproduced in
 Fig. (\ref{fig:ImSigma}) is from the energy dependence of the momentum distribution curves, which are also shown in the
 figure for various energies. Fits to the energy distribution curves for fixed momenta, also done in the same paper \cite{KaminskiPR2005} 
 gives results consistent with the parameters deduced. The  momentum distribution function, 
fits very well a Lorentzian. The Lorentzian form is evidence that the relaxation rate is also independent
of momentum perpendicular to the Fermi-surface \cite{AbrahamsV-PNAS}.
To convert to the scattering rate as a function of energy, one must multiply
 by the band-structure velocity at the measured $\omega$. 
 The low frequency departure from linearity as a function of $\omega$ is due to impurity 
 scattering and finite temperature. The data also deduces a frequency (and temperature) independent relaxation rate
 which is quite angle dependent. It is not possible in this experiment to disentangle the small angle 
 impurity scattering contribution 
\cite{AbrahamsV-PNAS} and the
 angle-dependent width due to bi-layer splitting in this quantity.
 
 \begin{figure}
\includegraphics[width=1.0\columnwidth]{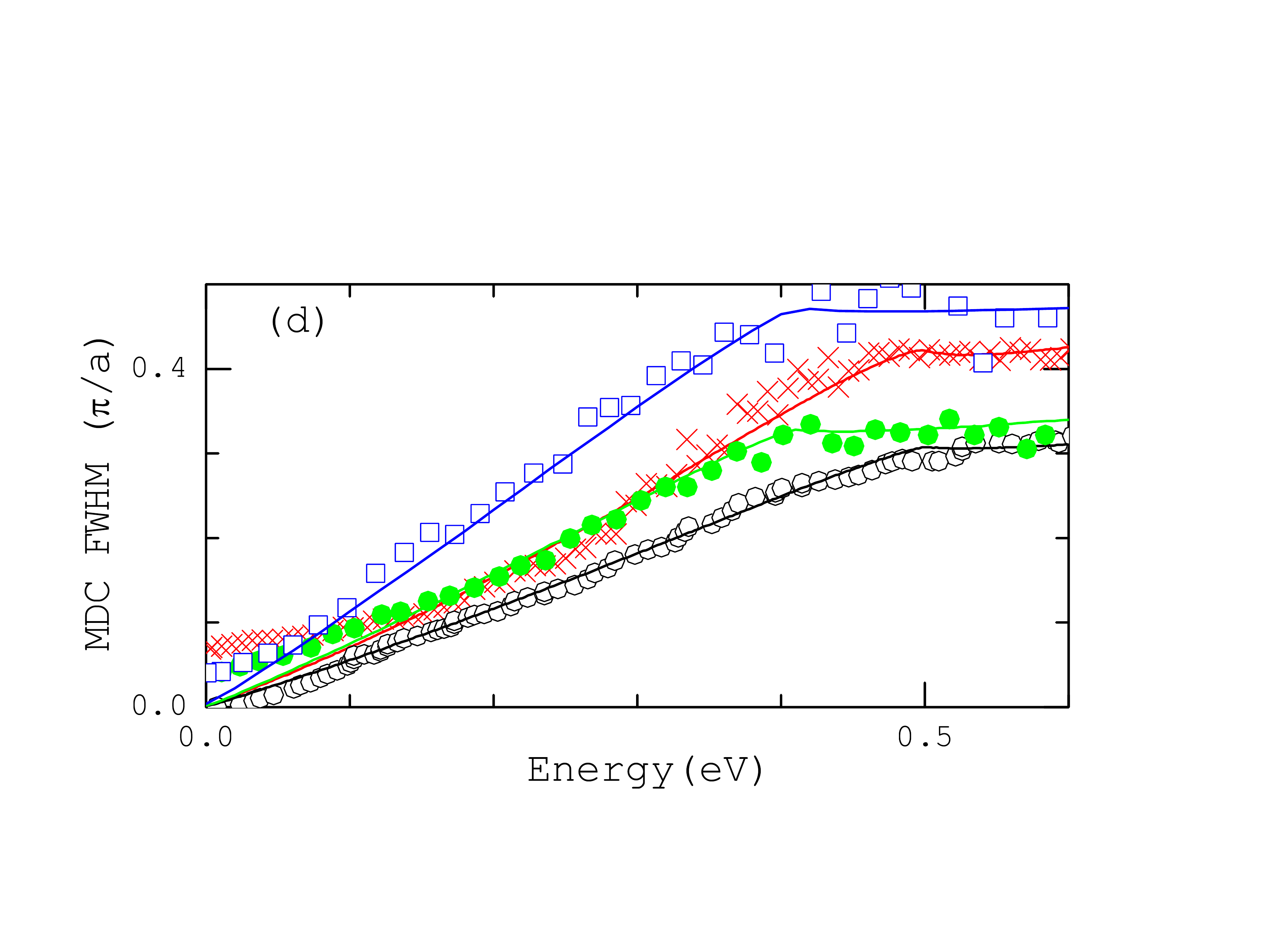}
\caption{The full width of the Lorentizian momentum distribution curves as a function of energy 
up to high energy for the three compounds which are measurable by ARPES.
The open black circles are data for optimally doped Bi2201 (nodal cut, from \cite{Meevasana})
The
red-crosses are for "optimally doped" Bi2212 (nodal cut, from \cite{Lanzara2007}), 
the full green circles are for $La_{2-p}Sr_pCuO_4$ 
for $p=0.17$ at about 20 degrees from the nodal direction and the blue
squares are for the same compound at $p=0.145$ from the nodal direction. 
The last two set of points are from
\cite{Chang2007}}
 \label{fig:ImDK}
\end{figure}

 The parameter $b = 0.7 \pm 0.1$ shown in fig. (\ref{fig:ImSigma}) is independent within this error bar of the momentum
 along the Fermi-surface. It is defined through $Im \Sigma = a +b \omega$ in the legend in the figure. Given the definition
  of the parameter
 $g_{\hat{p}}$ in Eq. (\ref{z}), $b = \frac{\pi}{2} g_{\hat{p}}$. This experiment therefore determines
 ${\overline{g}} = 0.4 \pm 0.1$. This should be compared with ${\overline{g}}$ deduced from the specific heat
 in Eq. (\ref{g,Tx}). It must be remembered that the specific heat is measured in a different compound than the
 scattering rates. However, when a variety of compounds are measured, as in the resistivity results \cite{Taillefer-Planck} 
 quoted below, the variation
 of this parameter appears to be no more than 50\%. The measurements of the single-particle line
  shapes in $La_{2-x}Sr_{x}CuO_4$ \cite{Chang2007}
\cite{ZhuVPRL2008} show angle dependence in the scattering rate increasing by 
about 50\% from the $(\pi,\pi)$ to the $(\pi,0)$
directions with a value in the former about the same as that shown in Fig. (\ref{fig:ImSigma}).
 
 Fig. (\ref{fig:ImDK}) shows a compilation of data of the line-width in the momentum distribution for
 a fixed frequency up to frequencies of about 0.5 eV in three different
 compounds that are measurable by ARPES. 
 The low frequency departure from linearity as a function of $\omega$ is due to impurity 
 scattering and finite temperature.  The upper cut-off in the relaxation rate given by the theory, 
 when it must change to frequency independent, 
 is within 50\%  of the $\pi T_x$
  in the specific heat data shown in Fig. (\ref{fig:QcrSpHt}). This is no more than a  requirement of causality. 
   Later measurements in Bi2212 show \cite{BokPRB2010} that this upper cut-off is angle 
 dependent - about 0.5 eV near the diagonal and decreasing to about 0.2 eV towards the $(\pi,0)$ -directions.
  Below about 0.4 eV, the 
 cut-off is the same as the bottom of the band measured from the chemical potential \cite{KaminskiPR2005}.  
 
 \subsection{Resistivity}

\begin{figure}
\includegraphics[width=1.0\columnwidth]{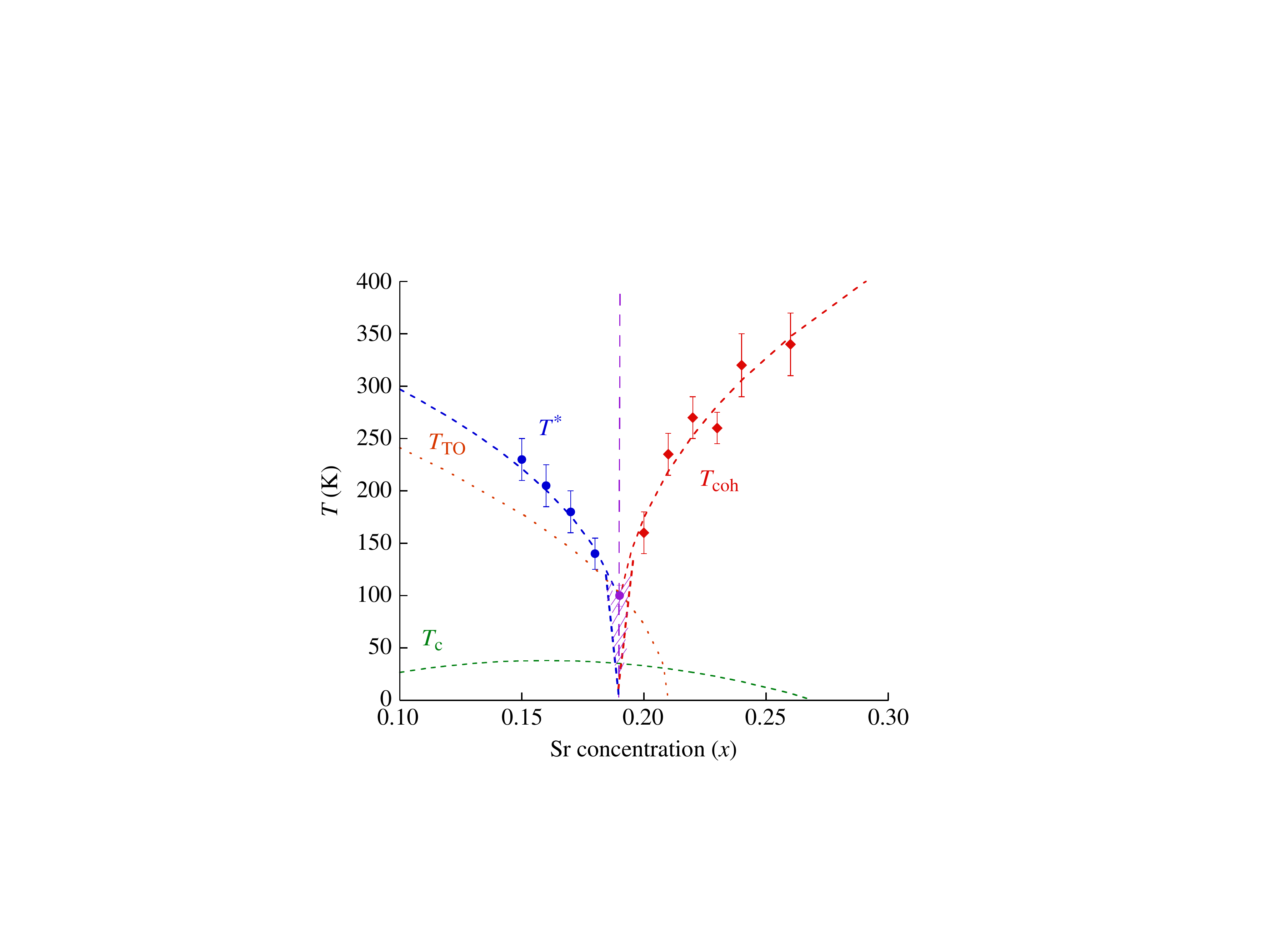}
\caption{The resistivity "phase diagram" from \cite{Hussey2011} for $La_{2-x}Sr_xCuO_4$. The temperature
dependence of the resistivity begins to show departure from linearity below the lines marked $T^*(x)$ and $T_{coh}$.
In this paper we are concerned with the latter. This line may be fitted with $(x-x_c)^{0.5}$, just as the cross-over 
line in the specific heat data above. More data between $x=0.18$ and $0.20$ would be desirable}
 \label{fig:ResPhDia}
\end{figure}

 Fig. (\ref{fig:ResPhDia}) presents the region of temperature as a function of doping $p$ in $La_{2-p}Sr_{p}CuO_4$
 in which the resistivity is linear in $T$ \cite{Hussey2011}. The dashed lines 
 give the temperature below which resistivity begins to
 deviate from linear in $T$.
 So the dashed line in the right marks the temperature cut-off $\xi_T^{-1}(p-p_c)$.
 
  The data is consistent with linear in $T$ resistivity to arbitrary low temperatures at near the critical 
  doping and in many compounds remains the same up to temperatures at which they begin to melt
  or decompose, about 1000 K.
Recently data from several compounds has been collected  \cite{Taillefer-Planck} and summarized after
a careful estimate of parameters such as electron density and velocity in terms of 
a transport relaxation rate
\be
\tau_{tr}^{-1} = \alpha k_BT/h, ~~for~~ p \approx p_c
\ee
I will identify 
\be
\alpha \equiv \frac{\pi}{2} ~g_{tr}.
\ee
where $g_{tr} \propto g$; their relation is discussed in Sec. III.
There is one fault in the deduction of the dimensionless parameter $\alpha$
in Ref. \cite{Taillefer-Planck} which is otherwise a very useful and careful paper.
The effective mass in the formula for conductivity
should be the band-structure mass and not the renormalized many body mass, which occurs for example 
in
the specific heat. This follows from a Ward identity \cite{Nozieres-book}, \cite{Varma85-HF}, \cite{Kadanoff-Prange},
 \cite{MIYAKE1989} 
which is a consequence of the continuity equation. This is a subtle point
which is dealt with in Sec.III. Here I simply note that 
if a renormalized mass were to be used, which is 
logarithmically divergent at criticality at low temperatures, the resistivity would not be linear in temperature but
be proportional to $T \ln T$ in the quantum critical region, which can be excluded. Moreover even if only a constant mass enhancement factor
 is put in as done in Ref. \cite{Taillefer-Planck},
 the momentum transport scattering rate
would be significantly larger than the single-particle scattering rate. The general theorem 
is that the transport scattering rates must always be smaller or equal to 
the single-particle
scattering rates, as discussed further below in Section III. Also as explained later, the Kadawaki-Woods relation \cite{KADOWAKI1986}
between the resistivity and specific heat of heavy-fermions and others $\rho(T) \propto T^2$ would not be followed. Instead it would be
$\rho(T) \propto T^3$.
The correct estimate of $\alpha$ is therefore about a factor of 3 
(the effective many body enhancement estimated in 
\cite{Taillefer-Planck}) lower than that given in that paper.  Ref. \cite{Taillefer-Planck} finds
$\alpha$ varying in different cuprate compounds to be 0.7 $\pm 0.2$ to 1.2 $\pm$ 0.3.  The corrected value then varies from
about $0.25  \pm 0.1$ to about $0.4 \pm 0.1$.   $g_{tr}$ is then about $2/3$ of these numbers.
$g_{tr}$ is in fact calculated to be about $(2/3)~ \overline{g}$ in Sec III,
 quite consistent with the single-particle scattering rates 
and the specific heat.
The resistivity phase diagram Fig. (\ref{fig:ResPhDia}) should be compared with Fig. (\ref{fig:QcrSpHt}) 
for  specific heat near criticality. More data near the critical point would be nice to have. 
Given what we have, one may
 deduce similar value of the cross-over exponent $\zeta \approx 0.5$ from this plot as well. 
\subsubsection{Minimum scattering length}
As already mentioned, the single-particle scattering rate gives an upper limit to the transport scattering rate
or the inverse width of the momentum distribution function gives a minimum limit to the transport
scattering length $\ell_{tr}$. The maximum in the width of the momentum distribution may be read from
Fig. (\ref{fig:ImDK}). It is about 0.4 $(\pi/a)$ at an energy of about 0.4 eV (corresponding to a temperature
$\omega_{cx}/\pi$, of about 1600 K). The single particle mean-free path $\ell$ is the half-width and $k_F$ is about $0.8 \pi/a$ near critical
doping. The fears that the transport mean- free path obtained from resistivity $\ell{tr}$ is such that $k_F \ell_{tr}$ or $\ell_{tr}/a$ is smaller than 1, 
the so-called "Mott-Ioffe-Regel
limit" are therefore unfounded. (If the resistivity per 2-d conducting layer in the cuprates is written as 
$\frac{h}{2e^2} \frac{1}{k_F \ell_{tr}}$, the estimated $\ell_{tr}$ \cite{Taillefer-Planck} is nearly the same as $\ell$ for the same compound.)
We appear to be a factor of about 5 on the safe side of it.
The basis of the Ioffe-Regel limit  is the uncertainty principle used in deriving it at low temperatures; its use at high temperatures 
needs study.

\subsection{Density correlations}
 \begin{figure}
\includegraphics[width=1.0\columnwidth]{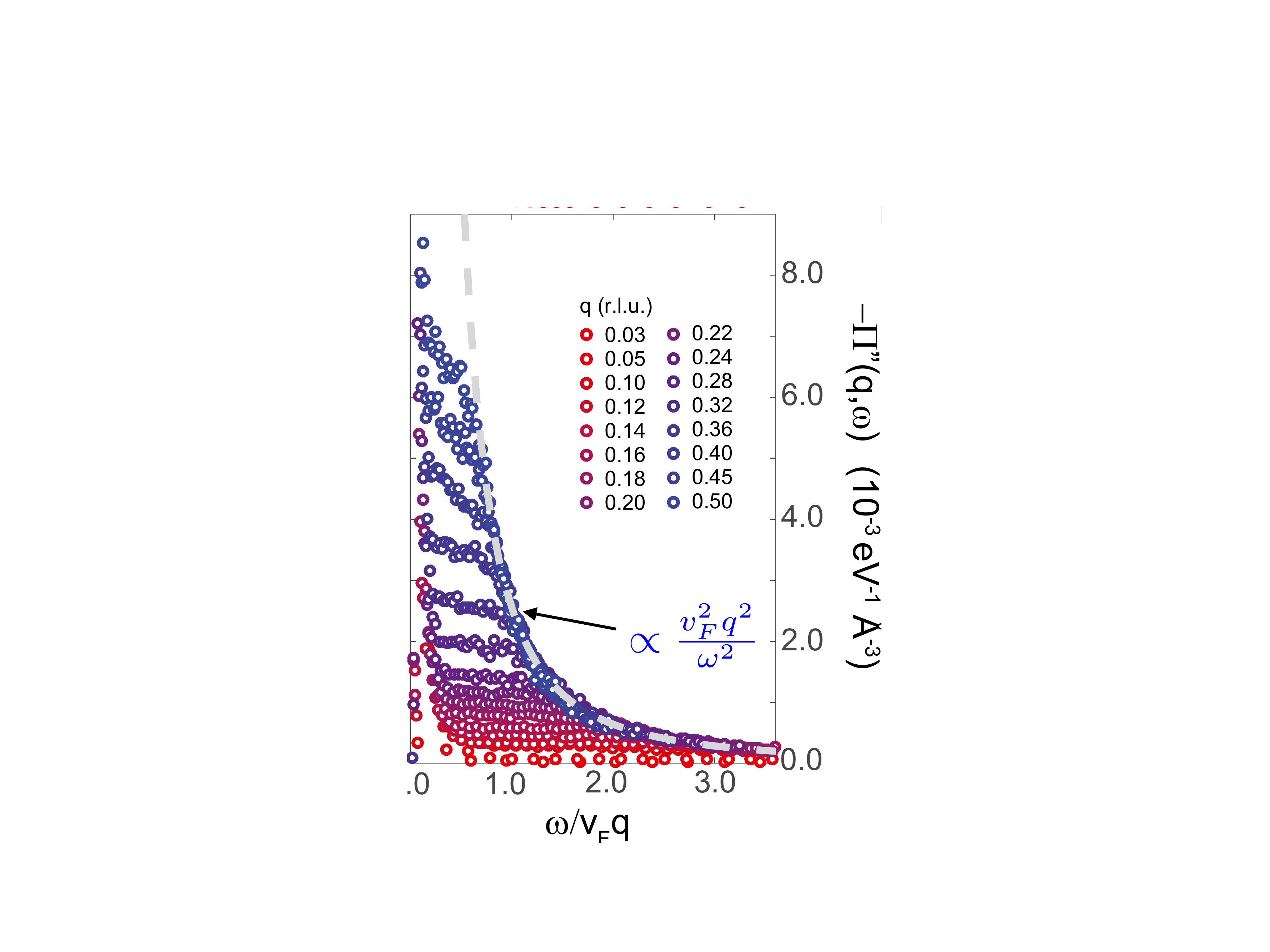}
\caption{The imaginary part of the "neutral" density-density correlation function taken from \cite{Abbamonte2018}
for the region $v_Fq/\omega \lesssim 1$, showing the fit to the square of this quantity whose coefficient is related to the
compressibility and the scattering rate using the
Einstein relation} 
\label{fig:DenCorr}
\end{figure}
Great technical developments have led to a laboratory instrument \cite{Abbamonte2018} to
 measure the density 
correlations accurately over a wide range of frequencies and over the entire Brillouin zone. 
These are shown in Fig. (\ref{fig:DenCorr}). Quite generally, the Einstein relation gives
 the conductivity
\be
\sigma(\omega, T) = e^2\kappa(T) D(\omega, T).
\ee
 $\kappa$ is the compressibility which is equal to the density of states at 
the Fermi-surface for non-interacting fermions for $T << E_F$. $D(\omega, T)$ is the diffusion function. 
Continuity equation gives that 
 the imaginary part of the (screened) density correlation function $\Pi"(q, \omega)$ for 
 $v_F q << \omega$,
\be
\label{chi}
\Pi"(q, \omega) =  \frac{ \kappa q^2 D(\omega)}{\omega}.
\ee
The possible quantum-critical aspects arise in the possible renormalization
of $\kappa$ and the frequency dependence of the diffusion function $D(\omega)$.
One can write 
\be
D(\omega) = \frac{v_F^2}{2} \tau_{tr}(\omega),
\ee
again with no renormalization in $v_F$ from its band-structure value and $\tau_{tr}$ is the transport relaxation rate.
 (\ref{chi}) is appropriate when
the velocity is isotropic. When it is anisotropic 
an appropriate average is called for which takes into account the direction of measurement of the density correlations.

Given $\tau_{tr}(\omega) \propto \omega^{-1}$,  
$\Pi"(q, \omega) \propto   \frac{q^2}{\omega^2}$ follows.  This is consistent with the optical conductivity
 if the conductivity is $\propto 1/\omega$ 
in the range of the data. (Actually both the logarithmic dependence of the mass
and the upper cut-off begin to play a visible role in the optical conductivity above a frequency 
of about 0.1 eV, but 
within the accuracy of the density correlation function data, they are unimportant.)

To compare quantitatively, the experimental results shown in Fig. (\ref{fig:DenCorr}) are fitted to find,
\be
\label{densSmallq}
\Pi"(q, \omega) = (3 \pm 0.5) \times 10^{-3} eV^{-1} \AA^{-3} \big(\frac{v_Fq}{\omega}\big)^2.
\ee
A bare fermi-velocity of about 2eV~$\AA$ obtained from ARPES measurements is used to get this result. 
The numerical coefficient then is equal to $\kappa (\pi/2) g_{tr}/2$. The theory of the density correlations in the
limit $v_Fq/\omega <<1$ \cite{Shekhter-V-Hydro} summarized below
allows no singular corrections to the compressibility but Fermi-liquid corrections are allowed.With 
the dimensionless $g_{tr}$ from the resistivity measurements of about 0.3, we get $\kappa \approx 10^{-2}/(eV\AA^3)$.
An unrenormalized $\kappa$ is the density of states near the chemical potential. 
Such a density of states is about $2 ~states ~/[2eV (16 \times 12)\AA^3] \approx (1/2)10^{-2}/(eV\AA^3)$.

Actually (\ref{chi}) is obeyed up to nearly $v_F q = \omega$ below which it is nearly a constant. 
The constant part is indeed even more remarkable than the part discussed above.
A theory for that \cite{Varma2017DenCorr} has also been provided. 

\subsection{Symmetry of Superconductivity in cuprates and parameters determining $T_c$}

Superconductivity of d-wave symmetry is observed in the cuprates. d-wave pairing requires
 that scattering of fermions on the fermi-surface be predominantly through $\pm \pi/2$ \cite{MiyakeSV1986}. It is axiomatic that the same fluctuations which 
dominantly scatter fermions in the normal state
are responsible for pairing the fermions. 
This poses a paradox in cuprates because as we see in Fig. (\ref{fig:ImSigma}) the 
single-particle scattering rate is nearly isotropic. The solution of this paradox has been provided \cite{ASV2010} and will be 
summarized in Sec. III. 
Its experimental verification comes through analysis of ARPES experiments in the superconducting state \cite{Bok_ScienceADV} 
where the spectral function of the fluctuations determining both the superconductivity and the normal state
scattering rates were both determined as well as the symmetry of the coupling functions in the normal and the pairing channels
 and the parameters for both the spectral functions and the coupling functions.  The experiments were done
on a sample of Bi2212 with a $T_c$ of 90 K.  It is explained in \cite{Bok_ScienceADV}  
that the flat frequency dependence of 
the theoretical and experimentally deduced 
 quantum-critical fluctuation spectra, 
 which is further described in Sec. III, leads to an enhancement for the effective 
 dimensionless parameters $\lambda_{s,d}$ for pairing  by a factor of 
 $\approx 3$ over $g$. This comes about because the 
 parameters $\lambda_{s,d}$, the $s$ and $d$-wave for interactions in the s-wave and d-wave pairing channels
 (given by Eq. (1) in \cite{Bok_ScienceADV}) are $g$ multiplied by an integral over all $\omega$ 
 of the appropriate angular averages respectively of the spectral weight of fluctuations divided by $\omega$.
  The spectrun is deduced to
 be essentially $\omega$-independent up to the upper cut-off $\omega_c \approx 0.25 eV$.
  So $\lambda_{s,d} \approx g \ln(\frac{\omega_c}{T_c})$. Therefore the 
  deduced $\lambda_{s,d} \approx 1.2$ corresponds to $g \approx 0.4$ for $T_c \approx 90 K$.
 
  The large value of the cut-off and the logarithmic  
  enhancement of coupling constant for pairing are crucial for the high $T_c$
  in the quantum-critical
 region.

\subsection{Resistivity and specific heat in Heavy Fermions and Fe based compounds}
\begin{figure}
\includegraphics[width=1.0\columnwidth]{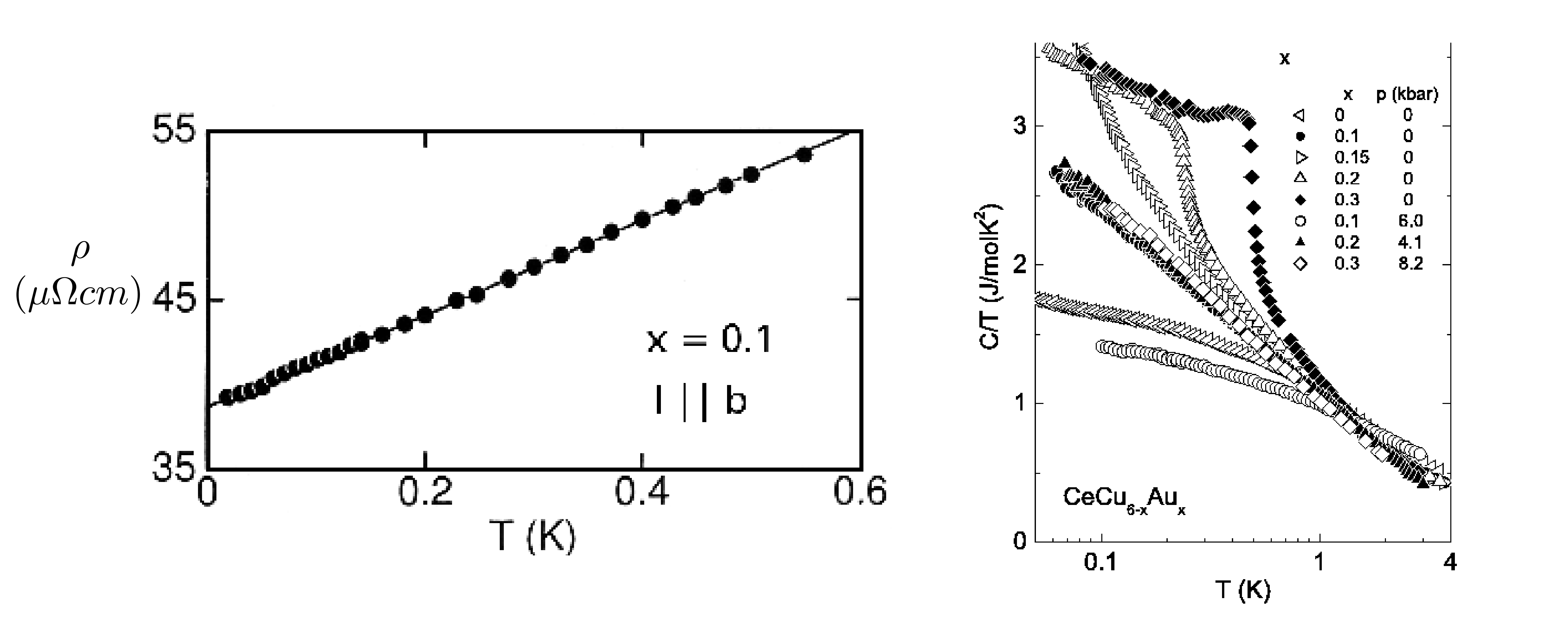}
\caption{Resistivity of $CeCu_{5.9}Au_{0.1}$ and specific heat at various pressures and dopings of
$CeCu_6$ across the antiferromagnetic quantum critical point. 
The figures are taken from Ref. \cite{HvL1996}.} 
\label{fig:RCCE}
\end{figure}

Temperature dependent resistivity proportional to T is measured from 30 mK to about 0.6 K 
in the AFM quantum-critical region of the compound CeCu$_{6-x}$Au$_x$ at $x = 0.1$
 with cross- over on both sides \cite{HvLRMP2007}. Correspondingly, the specific heat follows Eq. (1) 
 with cross-over in either side. See Fig. (\ref{fig:RCCE}). From the specific heat, one 
deduces $g \approx 0.8$, and $T_x \approx 10 K$. Using an effective Fermi-energy of about 20 K, 
corresponding to the background specific heat in the nearby Fermi-liquid compositions, 
a slope in resistivity of about 0.5 is obtained. The cut-off $T_x$ is 
similar to what is directly deduced from the measurements of the fluctuation spectra
 \cite{Schroder1}, \cite{SchroderZhuV2015}. 
 The energy scales in this compound are too small to be measurable in single-particle spectra by ARPES.
\begin{figure}
\includegraphics[width=0.8\columnwidth]{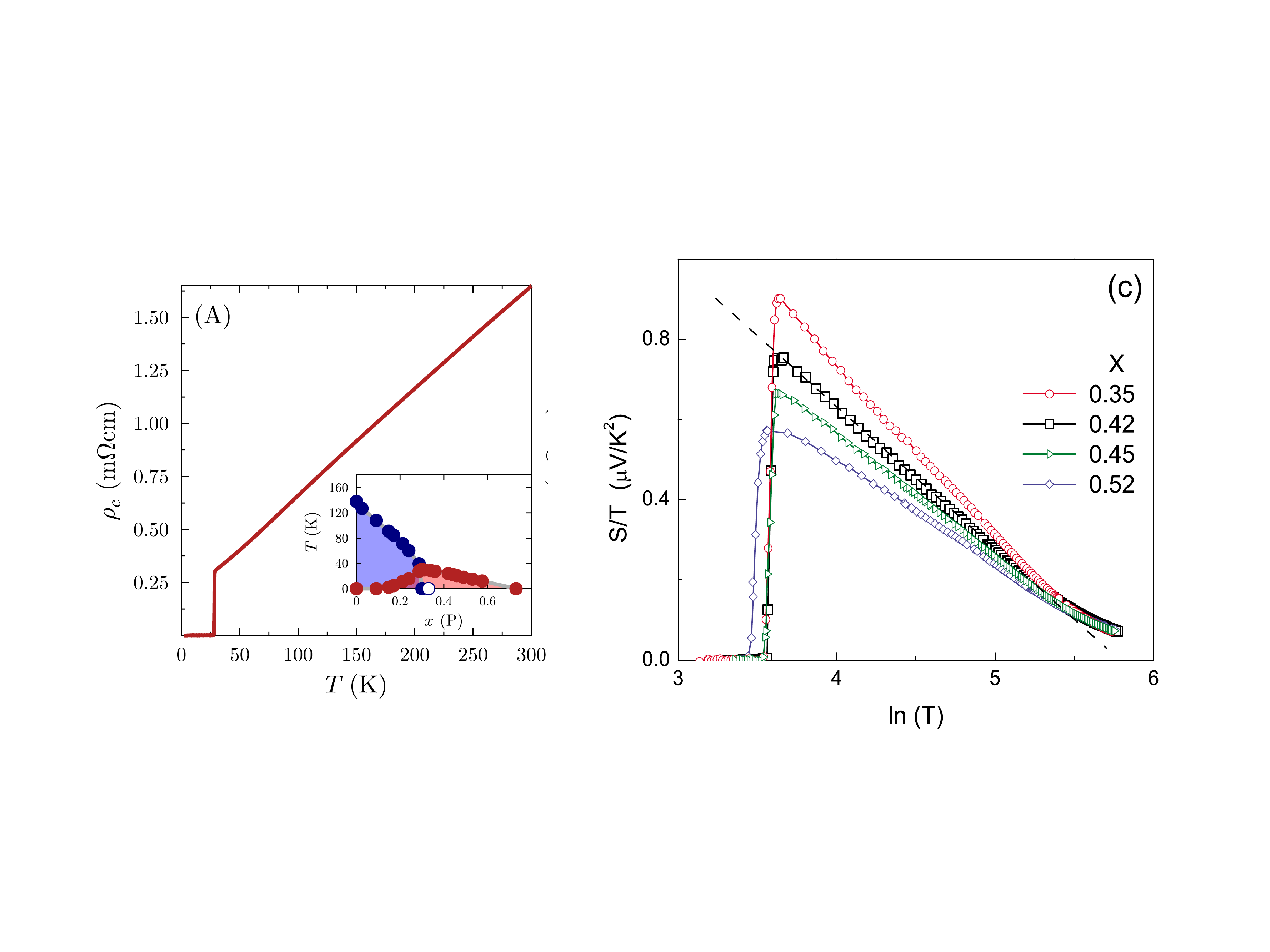}
\caption{Left: Phase diagram of $BaFe_2(As_{1-x}P_x)_2$ and Resistivity at criticality $(x\approx 0.3)$. 
Taken from \cite{Analytis2018}.  Thermopower S divided by T across AFM quantum criticality in $K_xSr_{1-x}Fe_2As_2$,
effectively showing the $T \ln T$ dependence of the specific heat at quantum-criticality. From Ref. \cite{thermopowerFeAS}.
The same paper also gives resistivity showing $\propto T$ behaviour in a similar range in temperature and composition.
 } 
\label{fig:R_S_Fe}
\end{figure}

In the Fe based compounds, evidence for the existence of a quantum critical point 
\cite{ShibauchiQCP} has been noted through linear in $T$ dependence of the resistivity. Resistivity
for one of the compounds showing one of the clearest linear in $T$ dependence is shown in 
(\ref{fig:R_S_Fe}) together
with the thermopower of another which also shows such a resistivity.
There is no single-particle spectra to compare with and neither are the basic band-parameters,
 an average fermi-velocity etc. known, or a background Fermi-liquid specific heat estimated to 
 get enough parameters to determine $g$ from a scattering rate. 
  Thermopower, which
  is the entropy per thermally excited particle does vary as as $T \log T$,
   consistent with the theoretical point of view in this paper. 
   But a quantitative estimate of the parameter $g$ from its magnitude 
   is not possible since there is no estimate of $\gamma$.
    These and the single-particle scattering rates will hopefully be available in the future.

\section{Foundation of the results above in a microscopic theory}

The microscopic theory has been briefly summarized \cite{Varma_IOPrev2016} and detailed references given. 
Only the motivations for deciding what is the relevant model to solve, and the principal results from its solution with direct 
applications to the experiments
discussed above are discussed below. 
\subsubsection{Order Parameter}
By the mid-1990's large and universal changes in thermodynamic and transport properties below a  line 
  $T^*(p)$ in the phase diagram of the cuprates were observed. 
The fact 
that the region of the  quantum fluctuations proposed phenomenologically \cite{CMV-MFL} abutted $T^*(p)$, suggested that it was a 
line of phase transitions to a broken symmetry, 
which terminated at the quantum-critical point when $p \to p_c$.  The order parameter, though it is required to have a condensation energy
typically larger than the maximum superconducting condensation energy, has to be unusual so that it was hidden in the experiments
carried out hitherto. An order parameter was suggested based on mean-field calculations on the three-orbital model for the cuprates \cite{VSA1987, Emery1987} 
which 
appeared to satisfy these
requirements. The order parameter is depicted on the left in Fig. (\ref{order}). It preserves  translation symmetry, is time reversal and 
and inversion odd but preserves their product. Also it is odd in three of the four reflection symmetries of the square lattice. 
 It can be algebraically represented by the magneto-electric or
anapole vector shown 
in Fig. (\ref{order}):
\be
\label{order}
{\bf \Omega} = \int_{cell} d^2r ~\big({\bf M(r)} \times {\bf \hat{r}}\big).
\ee
The magnetization ${\bf M(r)}$ is due to a pair of current loops in each unit-cell as shown in the figure. 
 \begin{figure}
\includegraphics[width=0.8\columnwidth]{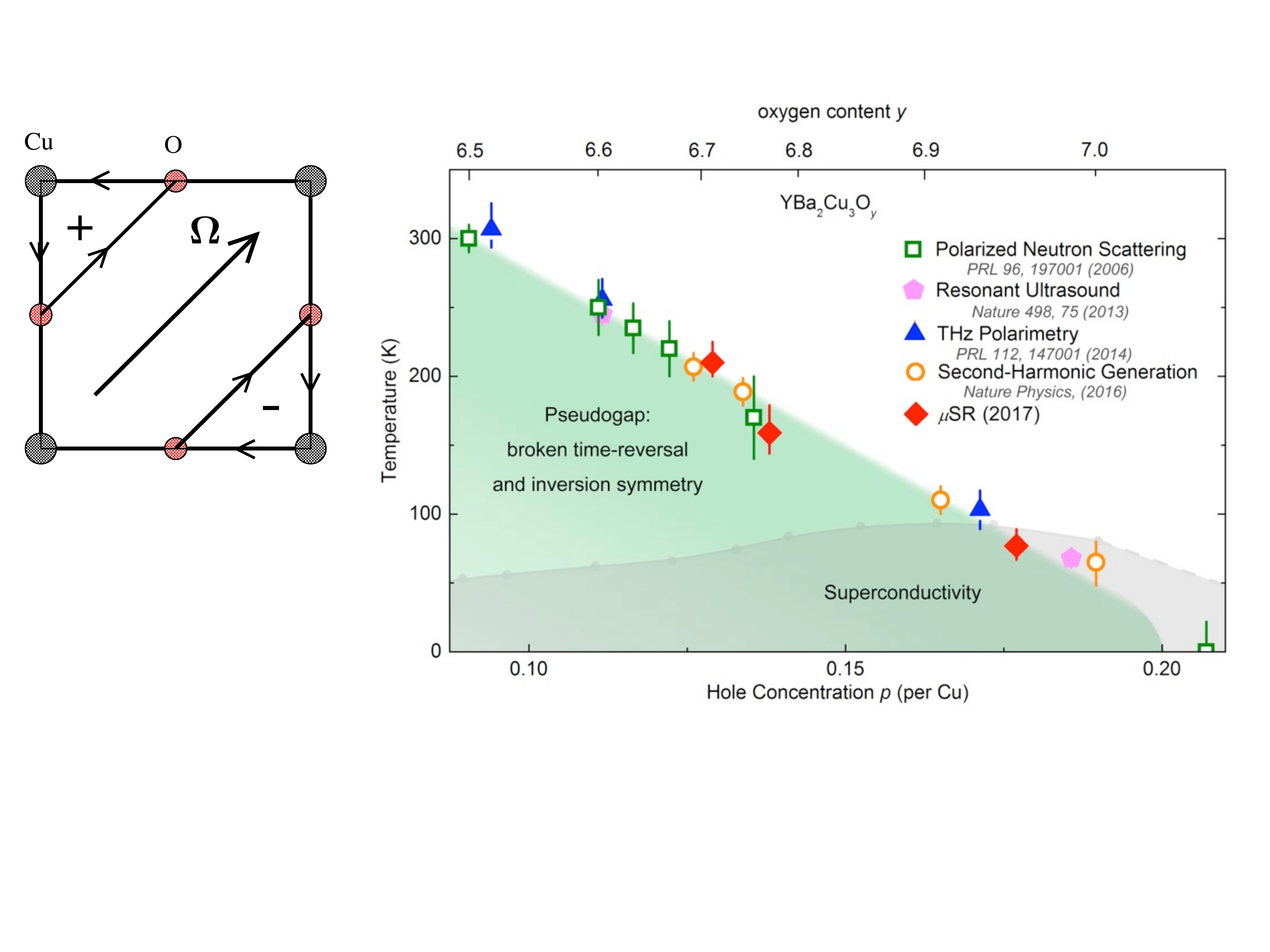}
\caption{Left: The order parameter depicted by the vector ${\bf \Omega}$ representing the magneto-electric order parameter of
Eq. (\ref{order}). ${\bf \Omega}$ is odd in both time-reversal and inversion and preserves their product. These symmetries come
from a pair of spontaneously generated current loops in a Cu-O$_2$ unit-cell. 
Right: Various experiments showing the onset temperature of symmetries consistent with the above in the compound YBaCu$_3$O$_{6+x}$. 
The neutron scattering
experiments are from \cite{Fauque2006}, the polarimetry experiments from \cite{Armitage-Biref}, the second harmonic generation from \cite{Hsieh2017}, 
the $\mu$SR from \cite{Shu2018}
and the ultrasound measurements from \cite{Shekhter2013}
} 
\label{fig:Order}
\end{figure}
The symmetries
 broken along the line $T^*(p)$ ending at the quantum-critical point are best observed
 by polarized neutron scattering \cite{Bourges-rev} although at least five different techniques have observed
 some or other of the broken symmetries predicted at the transition \cite{Kaminski-diARPES},
  \cite{Leridon}, \cite{Hsieh2017}, \cite{Kapitulnik1}, \cite{Armitage-Biref},  \cite{Shu2018}, \cite{Shekhter2013}, 
  \cite{Matsuda-torque1}, \cite{Matsuda-torque2}, \cite{Cerne2019}. 
 Neutron scattering experiments \cite{Bourges-rev} have observed changes consistent with the symmetries in Fig. (\ref{order})
for four different families of cuprates at various $p$. Fig. (\ref{order}) on the right shows the onset temperature of the symmetry changes observed 
in various experiments in the cuprate
most extensively investigated, $YBa_2Cu_3O_{6+x}$. The magnitude of the order parameter is typically 0.1 $\mu_B$ per unit cell. The free-energy reduction
due to such an order parameter setting in at 100 K \cite{CMV-Zhu-PNAS} is about three times that of the maximum superconducting condensation energy 
near $p_c$ fulfilling one of the
requirements that it be a transition competing with superconductivity and overcoming it at smaller doping.

\subsubsection{Model and correlation function for quantum- fluctuations}

The order parameter ${\bf \Omega}$, shown in Fig. (\ref{order}) has four possible orientations in a unit-cell. The interactions between cells is between these four
possible orientations. The model for the order parameter is therefore the 
 the two-dimensional xy-model with four fold anisotropy if one ignores the amplitude fluctuations which are irrelevant in two dimensions. 
  In the classical xy model, the four-fold anisotropy is marginally relevant.
But it is shown \cite{Aji-V-qcf1} to be irrelevant for a quantum phase transition. The classical model has no divergence of the specific heat,
 consistent with the lack of any sharp signature in the experimental specific heat at $T^*(p)$. 
 The model to be solved then is the quantum XY model in 2d, whose fluctuations are coupled to the fermions in the model. 
  An essential aspect of a quantum phase transition in a metal is the
 dissipation due to decay of the order parameter to incoherent degrees of freedom of the same symmetry in the fermions.
 So the model has the action:
 \be
\label{action1}
S &=&-K_0 \sum_{\langle {\bf x, x}' \rangle} \int_0^{\beta} d \tau \cos(\theta_{{\bf x}, \tau} - \theta_{{\bf x}', \tau}) 
 + \frac{1}{2E_{0}} \sum_{{\bf x}} \int_0^\beta d \tau \left( \frac{d \theta_{{\bf x}}}{d\tau}\right)^2  + S_{diss}.
 \ee
 $\theta_{{\bf x}, \tau}$ is the angle of the vector ${\bf \Omega}$ in the unit-cell at location ${\bf x}$ and at imaginary time $\tau$,
 which is periodic in the interval $(0, \beta = 1/k_BT)$. The first term is the potential energy of the  2d-XY model, the second is the 
 kinetic energy written in terms of the angular momentum ${\bf L}_z = \frac{d \theta_{{\bf x}}}{d\tau}$.  
 The third term is the dissipation
 due to coupling to fermions. The dissipation is of the Caldeira-Leggett \cite{CaldeiraLeggett} symmetry which was introduced for the dissipation of current fluctuations
 proportional to the gradient of the superconducting phase in a bath of current carrying fermions. In the present case, its origin is the decay of the collective current, 
proportional to $\nabla \theta({\bf r}, \tau)$ or that of the collective angular momentum variable ${\bf L}_z({\bf r}, \tau)$ into local angular momentum fluctuations 
of fermions. For further details, please see  \cite{Varma_IOPrev2016}. Each of the terms of this model has been derived from the three orbital model \cite{ASV2010}.
 
 The classical xy model was solved analytically by a renormalization group method \cite{Kosterlitz} after transforming it essentially exactly in to a model for vortices
   interacting logarithmically in space.  Remarkably, the quantum xy model with the action (\ref{action1}) can be transformed also essentially exactly \cite{Aji-V-qcf1} into a model
 for vortices interacting logarithmically in space  but locally in time even for the quantum model and another set of topological excitations, the {\it warps}, which interact 
 dominantly through a 
 logarithmic interaction in imaginary time $\tau$ but locally in space. This model can be solved also by the renormalization group method \cite{Hou-CMV-RG}.  
 The vortices and warps
  are orthogonal topological excitations which is what makes the
 model soluble. The answers are checked by
 detailed Monte-Carlo calculations \cite{ZhuChenCMV2015, ZhuHouV2016} on the original model (\ref{action1}), where evidence for the vortices and the 
 warps and their
 correlations is directly exhibited.
 
 The correlation function of the operator  $e^{i \theta({\bf r}, \tau)}$ is calculated 
 in the quantum-fluctuation regime of the model.
 The most important result is that the correlation functions are
 products of a function of space and of imaginary time $\tau$ and
  that the spatial correlation length is proportional to logarithm
 of the temporal correlation. This leads effectively to relative freedom
  of the temporal and spatial metric near criticality and 
 to very unusual but simple results for physical properties in
  terms of just two parameters.
  \be
 \label{corrfn}
 G({\bf r, r'}, \tau, \tau') &=& \langle e^{i \theta({\bf r}, \tau)}e^{-i \theta({\bf r'}, \tau')}\rangle \\ &= &G_0 \Big(\frac{\tau_c}{\tau-\tau'}\Big) \Big(\ln{|({\bf r-r'})/a|}\Big)
 e^{-|\tau-\tau'|/\xi_{\tau}} e^{-|{\bf{r-r}}|'|/\xi_{r}}; \\ 
\frac{\xi_{r}}{a} &=& \ln \frac{\xi_{\tau}}{\tau_c}.
\ee
$\tau_c^{-1}$ is the high energy cut-off in the theory ($\omega_{cx} = \pi T_{cx}$ in the analysis of
 the experiments.) The $\frac{1}{\tau-\tau'}$ dependence can be transformed to Matsubara frequencies which can be analytically transformed
 to real frequencies  to give the function 
 $\tanh(\omega/2T)$. The asymptotic low energy and high energy forms of this were the
 phenomenological assumption made in 1989 \cite{CMV-MFL} for the fluctuation spectra. It is also shown that the
 spectral fluctuations of the correlation $< {\bf L}_z({\bf r}, \tau) {\bf L}_z({\bf r'}, \tau'>)$ are identical to Eq. (\ref{corrfn}) \cite{Aji-V-qcf2}.

 The variation of the correlation length $\xi_{\tau}$ as a function of  $(p -p_c)$ or equivalently 
to the parameters of the xy model has been obtained by quantum Monte-carlo calculations 
\cite{ZhuHouV2016}. When the variation is due to the variation in the ratio of the kinetic energy
 parameter to the interaction energy parameter for a fixed dissipation, 
 the exponent $\zeta$ defined through the experiments on specific heat and resistivity above,
  is approximately 1/2,  consistent with the data.

Just like for the classical xy model,  quantum-critical fluctuations of the model are quite unlike the 
extensions of classical dynamical critical phenomena of models of 
the Ginzburg-Landau-Wilson class to the quantum regime. This appears to be essential as the conventional
critical dynamics of such a class cannot give the observed properties. In such models, the space and time are
correlations are connected through a finite dynamical critical exponents and the scale of both of the fluctuations
diminishes to zero near the critical point. By contrast the temporal fluctuations in (\ref{corrfn}) remain on the scale of
the cut-off over the entire fluctuation regime and the low energy form have a $\omega/T$-scaling. These were
essential in the calculations of  the observed experimental properties summarized above as well as for high temperature superconductivity.

\subsubsection{Coupling  function of fermions to the fluctuations}

 The coupling function of the fermions to the quantum-critical fluctuations should be used to calculate both the normal self-energy as well as the pairing self-energy.
 It should also be the same function which is used to calculate the dissipation of the quantum-critical fluctuations due to decay into particle-hole pairs. 
 The {\it important} coupling of the fermions to the fluctuations is deduced as follows: As discussed above, the critical fluctuations are also the fluctuations of the
 collective angular momentum operator $L_z({\bf r}, \tau)$. This can couple to fermions only through their local angular momentum operator with symmetry that of 
 ${\bf \ell}_z  \equiv \frac{1}{2} i ({\bf r} \times \nabla{\bf r} -  \nabla{\bf r} \times {\bf r})$. 
 So a Hamiltonian for the scattering of fermion creation and annihilation operators $\psi^+({\bf p}, \sigma), \psi({\bf p}, \sigma)$ 
  to the fluctuations is derived \cite{ASV2010} to be
 \be
 \label{coup}
 H_{coupling} = \sum_{{\bf p,p'},\sigma} \gamma({\bf p,p'}) ~\psi^+_{{\bf p},\sigma}~ \psi_{{\bf p'},\sigma} ~{\bf L}_z({\bf p-p'}) + H.C.
 \ee
 where the coupling function 
 \be
 \gamma({\bf p,p'}) =  i~ \gamma_0 ~ \big({\bf p} \times {\bf p'})
 \ee 
is the Fourier transform of the fermion angular momentum ${\bf \ell}_z$.  (In Ref. \cite{ASV2010}, equivalent of Eq. (\ref{coup}) is derived microscopically for the lattice fermions.) 
Below, the physics of how (\ref{coup}) leads to d-wave superconductivity and nearly angle-independent normal self-energy
is shown.

\subsection{Some measurable properties}
 
 \subsubsection{Single-particle self-energy and specific heat}
  The fluctuations (\ref{corrfn}) serve as the irreducible vertex \cite{Nozieres-book} in a calculation of the properties of the fermions. Because 
 of the product form of (\ref{corrfn}), the calculations for the properties of the fermions can be done easily
  and precisely \cite{Varma_IOPrev2016}.
 Of relevance to this paper is the retarded self-energy of the fermions, which has the full symmetry of the lattice. Since the fluctuations are momentum independent,
 only the projection of  $|\gamma({\bf p,p}')|^2$ to the full symmetry of the lattice appears in the calculation \cite{ASV2010, Bok_ScienceADV} as explained below. 
 For a circular Fermi-surface, that is just identity. Then,
 \be
\label{self-en}
\Sigma({\bf p}, \omega) &=&  g_{{\hat{\bf p}}} \Big(i  \frac{\pi}{2} ~ max(|\omega|, \pi T) + \omega \ln(\frac{\omega_{cx}}{x})\Big) , ~ for~ max(\omega, \pi T) \lesssim \omega_{cx}, \\ \nonumber
&= &  i \frac{\pi}{2} ~ g_{\hat{{\bf p}}} ~ \omega_c, ~\text{for} ~max(|\omega|, T) \gg \omega_{cx}.
\ee
$g_{\hat{{\bf p}}}$ is \cite{ASV2010} the product of  the amplitude of the fluctuations $G_0$, the density of states of the fermions $N(0)$ 
and the coupling $\gamma_0^2$. It also depends on the anisotropy of the band-structure. It is independent of ${\bf p}$ for a circular Fermi-surface. 
For the band-structure of Bi2212 near optimum doping, it is estimated to increase by less than about 2 
in going from the $(\pi,\pi)$ directions to the $(\pi,0)$ directions. 
For a square lattice, the correction to isotropy varies as $\cos 4 \theta({\hat{{\bf p}}})$.

The electronic specific heat can be written directly in terms of integrals over the imaginary part of the self-energy \cite{AGD}. 
The result is particularly simple when the self-energy is nearly momentum independent. 
Then the specific heat singularity is simply given by the inverse of the  renormalization given in Eq. (\ref{z}) which follows from
(\ref{self-en}). There is no renormalization of the compressibility if the self-energy is momentum independent \cite{Varma85-HF}.
\subsubsection{Resistivity}
The conductivity may be calculated in three different ways:\\
 (1) From the Kubo formula, which simplifies enormously
when the self-energy is independent of momentum \cite{CMV-MFL}. Then the
momentum transport scattering rate is equal to the imaginary part of the single-particle self energy. 
So the dc resistivity is proportional to $T$
in the quantum-critical region with the same slope as the self-energy for a circular Fermi-surface. 
One may calculate  the 
vertex correction due to the small angular dependence
in the self-energy using the Boltzmann equation, for example as
 in Refs. \cite{Varma-AbrahamsPRL}\cite{Abrahams-VarmaPRB}. 
 The transport scattering rate is then necessarily
 smaller than the single-particle scattering rate. A straightforward calculation shows that
with a factor of 2 variation in the self-energy, increasing in the direction
 where the velocity is least, 
the coefficient of the linear in T resistivity is about $2/3$ of that of the
maximum self-energy.\\
(2) The density-density correlation has also been calculated directly 
\cite{Shekhter-V-Hydro}, \cite{Varma2017DenCorr} giving 
the form shown in Fig. (\ref{fig:DenCorr}) with the inverse diffusion coefficient is related to the resistivity as noted already.
\begin{figure}
\includegraphics[width=0.8\columnwidth]{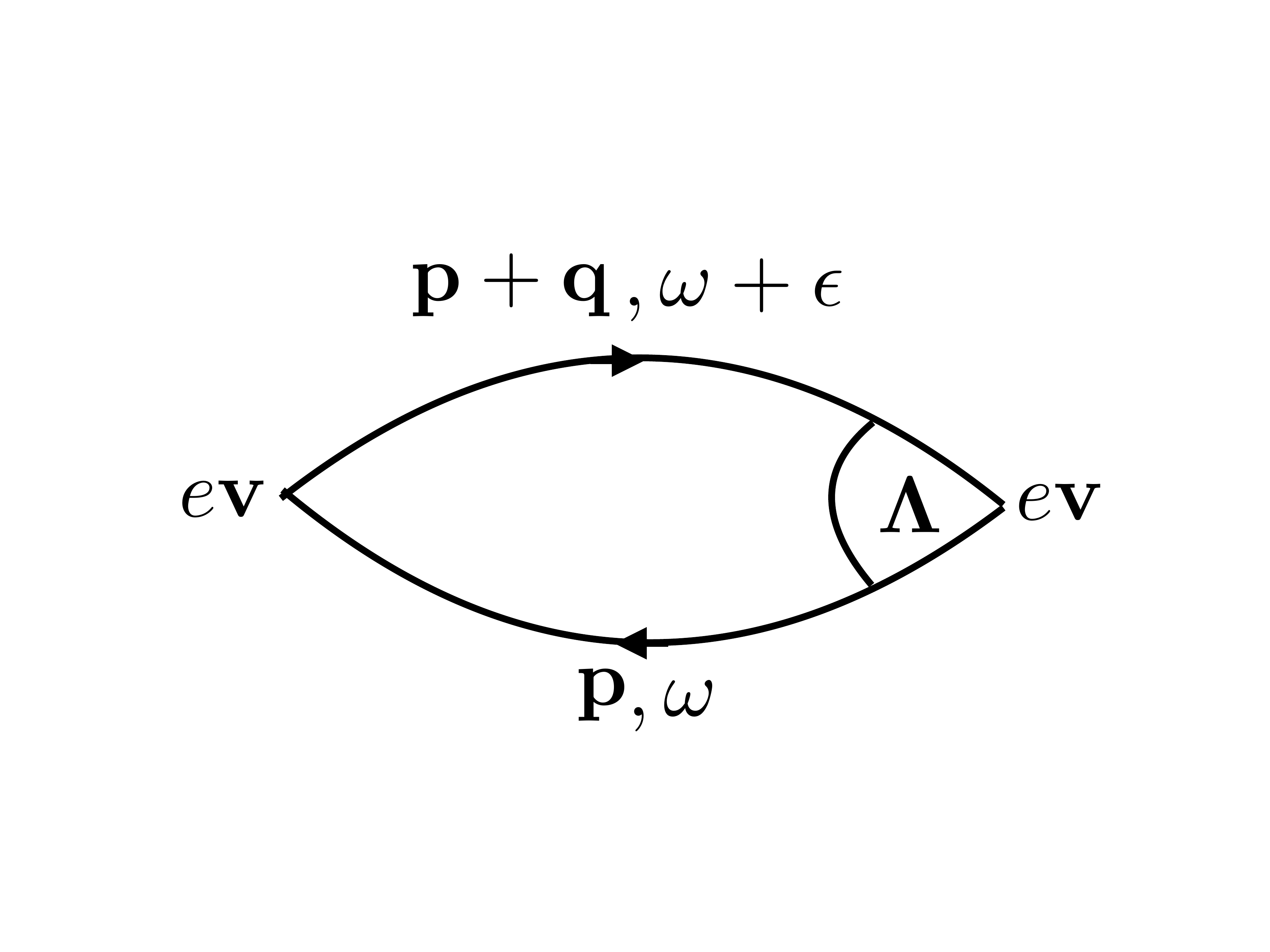}
\caption{The Kubo formula for the current-current correlation in terms of the bare current operator, the renormalized current operator
and the exact single-particle Green's functions. The dc conductivity is $(1/\epsilon)$ times the imaginary part of the current-current
correlation at ${\bf q} \to 0$.
 } 
\label{fig:cond}
\end{figure}

(3) From the Kubo formula for the conductivity: The Kubo formula for conductivity is equivalent to the evaluation of the diagram
 Fig. (\ref{fig:cond}) for the current-current 
correlations.
The vertex on the left is the bare (band-structure) velocity operator ${\bf v}$, the one on the right is the renormalized velocity
operator ${\bf v}_{renorm}$, and the lines are the exact Green's functions.  In the appropriate limit for calculation of dc conductivity, 
the matrix elements of the operator
${\bf v}_{renorm}$ are given by a Ward identity \cite{Nozieres-book} which follows from the equation for continuity:
\be
\label{ward-id}
{\bf v}_{renorm} &=& {\bf \Lambda}~ {\bf v} \\
Lim_{\epsilon \to 0{\bf q}\to 0} ~{\bf \Lambda}({\bf p},\omega;{\bf q},\epsilon) &=& {\bf 1} - 
\frac{1}{\bf v}}{\frac{\partial \Sigma({\bf p}, \omega, T)}{\partial {\bf p}}.
\ee
If $\Sigma({\bf p}, \epsilon,T)$ is independent of ${\bf p}$, the relevant limit of ${\bf v}_{renorm} = {\bf v}$. 
The sum over frequencies and integration over momentum over the
Green's functions in (\ref{fig:cond}), for the same conditions can be easily carried out. The  
 conductivity  in the $\alpha$ direction is then ,
\be
\label{sig}
\sigma_{\alpha}(T) = \frac{e^2 \langle v^2_\alpha\rangle N(0)}{2 Im \Sigma({\bf p}_F,0,T)}.
\ee
The dc resistivity is proportional to $T$ due to the self-energy in the denominator. 
 To get the coefficient of the resistivity, note that
 $N(0)$ in Eq. (\ref{sig}) is the bare (given by one-electron theory band-structure) 
 density of states and $<v^2_\alpha>$ is the bare 
 mean square Fermi-velocity in the 
$\alpha$-direction because of Eq. (\ref{ward-id}).  $N(0)<v^2_\alpha>$ is proportional 
to the inverse mass. It is for this quantity that often, as in Ref. \cite{Taillefer-Planck},
a renormalized mass, such as occurs in specific heat, is used.  As shown here, 
for the conditions of quantum criticality in cuprates, this mass 
is the bare or band-structure mass.
One can show \cite{Varma2017DenCorr} that the lack of any renormalization 
of the effective mass in the conductivity is true only
 for frequency much smaller than
the temperature. 

This issue arises in Fermi-liquids also if the condition that the self-energy is very weakly
 momentum-dependent compared to its
frequency dependence \cite{Varma85-HF} is satisfied. An application in that case to 
the resistivity in heavy fermions is given in Eq. (3) of Ref.  \cite{MIYAKE1989}, 
where the Kadowaki-Woods \cite{KADOWAKI1986}
observation was derived. 
For applications to the problems of interest in this paper, the $T^2$ on the left side of that 
equation should be replaced with $T$ and the $Im \Sigma$ on the right side 
by the relevant part of Eq. (\ref{self-en})to get Eq. (\ref{sig}) above. 
If one used a renormalized value for $N(0)<v^2_\alpha>$, resistivity would not
be proportional to the square of the specific heat (the Kadawoki-Woods observation) 
but be proportional to the cube of the specific heat.

For all the properties considered in this paper, there are only two parameters $g$ and $\omega_c$ in terms of which
every property considered is given. These two parameters were estimated in Ref. \cite{ASV2010} from the kinetic and interaction
energies of  a copper-oxide three orbital model. This gave $g \approx 1$ and
$\omega_c \approx 0.5 eV$, i.e. within about a factor of 2 of the experimentally deduced numbers.

\subsubsection{Coupling function for d-wave superconductivity}
Note that with ${\bf p,p'}$ on the (nearly) circular Fermi-surface
\be
\label{angle}
 -|\gamma({\bf p,p'})|^2 = -\gamma_0^2 |i \big({\bf p} \times {\bf p'})|^2 \propto \frac{\gamma_0^2}{2} \big(1 - \cos(2 \theta_p) \cos(2 \theta_{p'}) - \sin(2 \theta_p)\sin(2 \theta_{p'})\big).
 \ee
 Since in a calculation of the normal state self-energy, the intermediate state has the full symmetry of the lattice, see Fig. (\ref{fig:SEs}), and the fluctuation spectra is
  momentum-independent, the only angle dependence comes from the projection of the pair of vertices to the full symmetry of the lattice, i.e. of (\ref{angle}).  This yields just 
  $-|\gamma_0|^2$ and is repulsive in this symmetry for pairing. (The minus sign comes 
  from the loop integral in the diagram for the self-energies.) On the other hand in the pairing channel of d-wave symmetry, the intermediate state, has d-wave symmetry.
  So a projection of (\ref{angle}) to the d-wave can contribute from the second or third term in it. This is attractive and on a circular Fermi-surface would be 
  degenerate between the $d(x^2-y^2)$ state and $d(xy)$. In the cuprates, the density of states is least in the diagonal direction of the Brillouin zone, favoring
   thereby the former.
    \begin{figure}
\includegraphics[width=0.8\columnwidth]{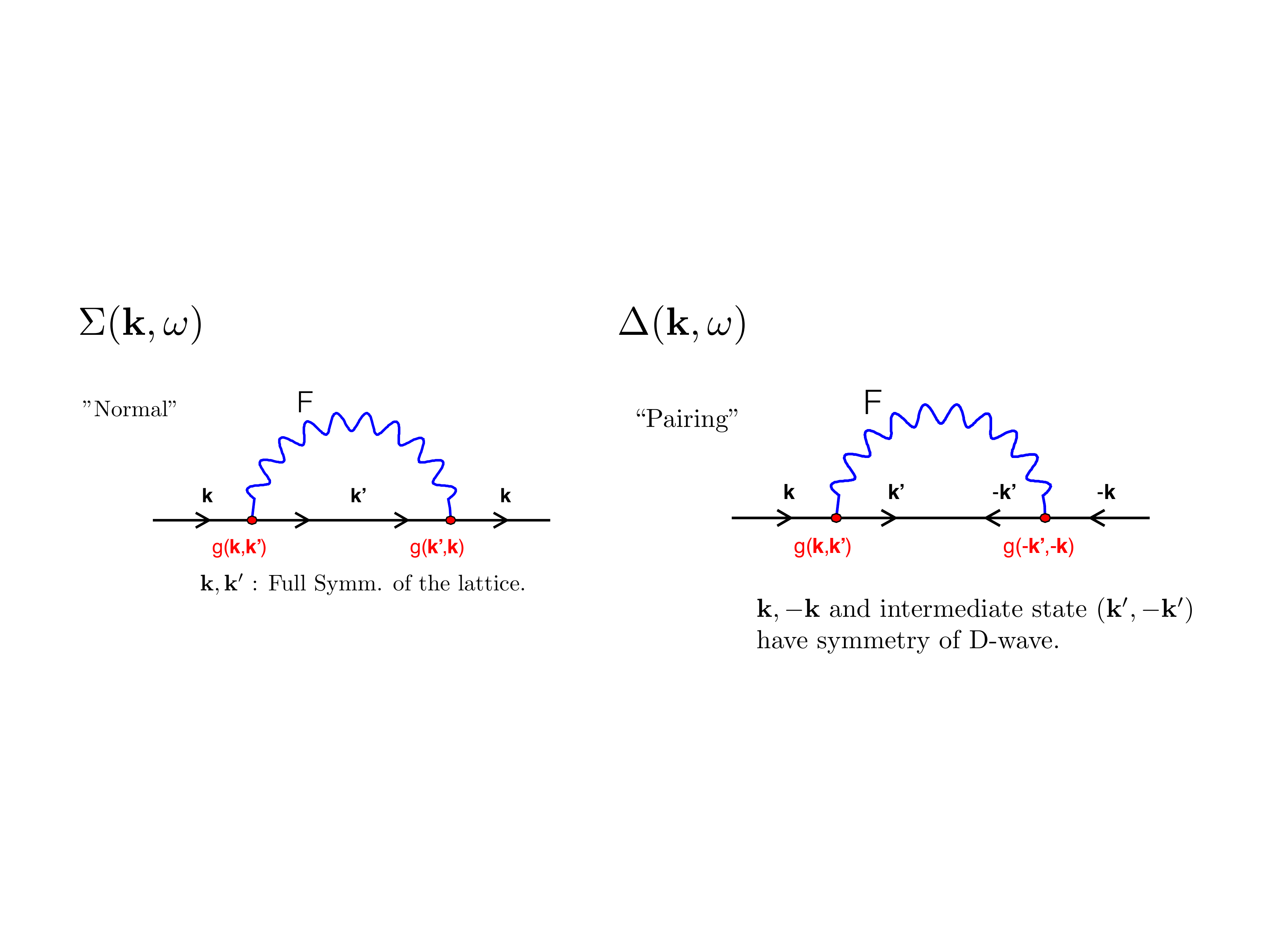}
\caption{The skeletal diagrams for the normal self-energy and the pairing self-energy  for scattering fermions from
 fluctuations whose propagator is depicted as a wavy line.  The former has the
ordinary part of the Gorkov Green's function in the intermediate state and the latter the pairing part. The vertices in both diagrams are closely related.
 } 
\label{fig:SEs}
\end{figure}
 
Eq. (\ref{angle}) and Fig.(\ref{fig:SEs}) explains the principal paradox in superconductivity of the cuprates:  While the measured 
 anomalous single-particle
 self-energy is nearly angle independent, superconductivity of d-wave symmetry occurs. 

\subsection{Related Matters}

No other theoretical ideas and calculations on any physical model have explained the temperature and frequency dependence of the
 properties discussed here in cuprates or heavy fermions or the Fe based  compounds, leave alone get the parameters or resolve the paradox
 of the symmetry of superconductivity discussed above.
 Calculations which are extensions of the dynamical classical critical phenomena to quantum-criticality 
 \cite{Hertz}, \cite{Moriya-book}, and extensively worked on since, are known not to give any of the experimental properties noted. 
 The closest that comes is a local model of $SU(N)$-spins in the limit $N \to \infty$
 \cite{SachdevYe} which 
 leads to the fluctuation spectra
 in frequency which was suggested in \cite{CMV-MFL}. Besides the difficulty of how such a model can be an
  effective physical model,
 it has the problem that an extensive ground state entropy is inevitably tied to its results. Modification of the model to get rid of this entropy
  \cite{Senthil2018} also alter its properties to that of a Fermi-liquid.

Some lacunae in the theory should be mentioned. \\
(1)The application of the
dissipative quantum xy model to the cuprates is rather clear given the symmetries changed  at $T^*(x)$. The
application of the same model is not a surprise close to the AFM critical point in the heavy fermions and the Fe compounds.
 The anisotropic AFM maps to such a model \cite{Varma2015afmqc} (see also erratum) and the
point of view is supported by the fit to the measured fluctuation spectra by the calculated fluctuations \cite{SchroderZhuV2015}.
The unanswered question is why the theory works over such a wide temperature range when the xy anisotropy
is so small that the classical transition would crossover to  that of the xy model only very close
 to the transition. The answer 
may lie in the possibility that with criticality of the kind discovered in the model, which for some purposes
may be regarded as having a dynamical critical exponent $z \to \infty$, the cross-over to
the anisotropic model occurs over essentially the entire range below the ultraviolet cut-off. The reason for the speculation is that
for the crossover temperature in classical critical phenomena is given in terms of the product $\nu z$, where $\nu$ is the classical correlation length
exponent. \\
(2) A principal problem in cuprates, the understanding of some remarkable properties in the pseduogap state, remain un-explained 
in the theory described above (and indeed in any other theory). While a
 thermodynamically significant phase transition of the predicted symmetry has been 
discovered at $T^*(p)$, it cannot give the peculiar observed Fermi-arcs \cite{ARPES-revZX} or the magneto-oscillations of a small 
Fermi-surface \cite{SebastianProust2015}, because
it does not change the translation symmetry of the lattice. Since the only phase transition discovered at $T^*(p)$ is to a phase
of the symmetry shown in Fig. (\ref{order}) and since its fluctuations explain so well both the quantum-critical region and the superconductivity
in the cuprates, a modification of the observed order in which a long period wave of the four distinct symmetries shown in Fig. (\ref{order})
has been suggested \cite{Varma-Per-Order-2019}. This retains the properties calculated and observed discussed in this paper and also is calculated
to lead to the unexplained properties \cite{Varma-Per-Order-2019}. The modified phase can in principle be discovered by high resolution resonant x-ray scattering.
Only after such an observation can one claim that the cuprate problem is solved.

  Recently, experiments in twisted bi-layer graphene reveal a linear in $T$ resistivity \cite{Herrero2019}, \cite{Efetov2019},
\cite{Young2019} in the phase diagram in 
a region
which has the quantum-critical shape in the phase diagram at the boundary between the insulator and
the superconductor, extending to asymptotic low temperatures when magnetic field is applied to suppress
superconductivity. One may speculate that the relevant model for critical fluctuations is again of the xy variety,
with the $U(1)$ symmetry being that of valley space,  which may be broken in the correlated insulator. 

I speculate that quantum-critical fluctuations of a variety of (soft) vertex 
models coupled to fermions, (many of which for 2D classical problems 
are treated in Baxter's book \cite{Baxter})
which classically are not in the Ginzburg- Landau, Wilson-Fisher class are generically related to  
the 2D- dissipative XY model in quantum version of the problems.  They may
all be governed by topological excitations with relative freedom of temporal and spatial fluctuations.
The scaling of the metric of space and time different from the flat world is the fundamental aspect of 
any quantum-critical problem. 
That the spatial and temporal metric become free relative to each other is a unique property which has led to
the extra-ordinarily simple results which explain the observed quantum-criticality in the problems discussed. 
The model itself is richer than the application noted here. 
For example, there is a critical region to a phase in it \cite{ZhuChenCMV2015, ZhuHouV2016, Hou-CMV-RG}
in which the correlation functions are of product form in space and time but with the temporal correlation length
 proportional logarithmically to the spatial correlation length. 
One may speculate, of course at great peril, that this is the appropriate description of another quantum-critical problem - 
inflation in the early universe.

{\it Acknowlwdgement}: I wish to thank the many many experimentalists who have  taken the time to explain their experimental 
results to me. The data quoted here is of-course a very tiny fraction of the mutually consistent data 
that exists in the literature. I especially wish to thank Bastien Michon and Louis Taillefer for providing me versions
of figures from their paper suitable for me and for answering questions about the data.
This paper was written while at Aspen Center for Physics.

\end{document}